\documentstyle[12pt]{article}
 \advance\oddsidemargin by -1in
 \advance\evensidemargin
 by -1in
 \marginparsep
 0.4cm
 \advance\topmargin by
 -0.0in
 \makeindex

 \newcommand\la{\langle}
 \newcommand\ra{\rangle}
 \newcommand\beq{\begin{equation}}
 
 \newcommand\eeq{\end{equation}}
 \newcommand\beqn{\begin{eqnarray}}
 \newcommand\eeqn{\end{eqnarray}}
 \newcommand{\doublespace} {
 \renewcommand{\baselinestretch} {1.6}
 \large\normalsize}

\begin{document}
\vspace*{3cm}

\begin{center}
 
{\Large \bf Broadening of Transverse Momentum}

\medskip
                                                                                
{\Large \bf of Partons Propagating through a Medium}
 
\vspace{1cm}
{\large M.B. Johnson$^{1}$, B.Z. Kopeliovich$^{2,3}$
and A.V. Tarasov$^{4,2,3}$}
\\[1cm]
$^{1}${\sl Los Alamos National Laboratory
Los Alamos, NM 87545, USA}\\[0.2cm]
$^{2}${\sl Max-Planck Institut f\"ur Kernphysik, Postfach
103980, 69029 Heidelberg, Germany}\\[0.2cm]
$^{3}${\sl 
Joint Institute for Nuclear Research, Dubna, 
141980 Moscow Region, Russia}\\[0.2cm]
$^{4}${\sl Institut f\"ur Theor. Physik der
Universit\"at, Philosophenweg 19,
69120 Heidelberg}
\end{center}

\vspace{1cm}
 
\begin{abstract}

Broadening of the transverse momentum of a parton propagating
through a medium                                           
 is treated using the color dipole formalism, which has the advantage of
being a well developed phenomenology in deep-inelastic scattering and soft
processes. Within this approach, nuclear broadening should be treated as
color filtering, {\it i.e.} absorption of large-size dipoles leading to
diminishing (enlarged) transverse separation (momentum). We also present a
more intuitive derivation based on the classic scattering theory of
Moli\`ere.  This derivation helps to understand the origin of the dipole
cross section, part of which comes from attenuation of the quark, while
another part is due to multiple interactions of the quark. It also
demonstrates that the lowest-order rescattering term provides an
$A$-dependence very different from the generally accepted $A^{1/3}$
behavior. The effect of broadening increases with energy, and we evaluate
it using different phenomenological models for the unintegrated gluon
density. Although the process is dominated by soft interactions, the
phenomenology we use is tested using hadronic cross section data.

\end{abstract}

\doublespace
 
\newpage

\section{Introduction} 

A high-energy parton propagating through a medium experiences
multiple interactions that increase its transverse
momentum. This broadening of the transverse momentum can be measured in 
different reactions, a few examples of which follow. 

In the Drell-Yan process \cite{e772}
the lepton pair produced carries undisturbed information
about the transverse momentum of the projectile quark which
undergoes initial state interactions.   The important condition is 
the shortness of the coherence time of the Drell-Yan process \cite{hir}, 
which allows one to factorize the initial 
state interaction from the cross section for the Drell-Yan reaction.
In the opposite regime of very long coherence time, the effect of
transverse momentum broadening also exists, although the lepton
pair is produced as a fluctuation long in advance the nucleus
(in the nucleus rest frame, see \cite{hir}) and does not ``communicate''
with the quark any more. Nevertheless, the nucleus supplies the fluctuation with
a larger
mean value of momentum transfer than a nucleon target, therefore it is
able to liberate harder fluctuations, {\it i.e.} those which have larger intrinsic
transverse momenta $k_T$. As a result, the value of $\la k_T^2\ra$ of the
lepton pair increases. The nuclear modification of the transverse momentum 
distribution of lepton pairs in the limit of long coherence time is calculated in
\cite{kst1,jkt}.

In a similar way, the production of heavy quarkonia off nuclei \cite{e772} 
can measure the $k_T$ broadening
of a projectile gluon. Final state elastic rescattering of
the produced quarkonium can be neglected since the cross section
is very small.

One can also study the $k_T$ broadening
of a quark originating from DIS on a nuclear target, which also includes the
Fermi motion of the participating nucleon. Although the
available data from the EMC experiment \cite{emc} do not show any
significant effect, this is related to a suppression by a factor $z^2$,
where $z$ is the fraction of the quark momentum carried by the produced 
hadron. In this experiment $\la z\ra \sim 0.25$. Nevertheless, new
precise data are expected soon from the HERMES experiment.

In the production of two high-$p_T$ back-to-back hadrons (or jets)
off nuclei, 
one of the hadrons defines the scattering plane, while the acoplanarity
of the other one serves as a measure for the nuclear broadening
of the transverse momenta by initial and final state interactions.
Available data \cite{fnal1,fnal2} demonstrate an unusually
strong effect (see the interpretation in \cite{k95}).

The propagation of a high-energy parton through a medium
can be treated intuitively as a random walk
in transverse momentum space leading to a linear increase of
$\la k_T^2\ra$ with the thickness of the matter covered
\cite{cp,hkp}. Such an interpretation faces, however, certain
difficulties. The cross section for the interaction of a colored quark
diverges at small momentum transfer, and one has to introduce an 
infrared cut off. On the other, hand the mean value of the momentum transfer
squared diverges at the ultraviolet limit.

It is demonstrated in Section~\ref{filter} that the broadening of the
transverse momentum of a parton propagating through a nucleus can be
treated within the light-cone dipole approach introduced in \cite{zkl} as a
color filtering effect \cite{bbgg} for a color $\bar qq$ dipole traveling
through nuclear matter. Color filtering is a size-dependent attenuation of
$\bar qq$ color dipoles propagating through nuclear matter.

Since the formal transition from a single quark to a $\bar qq$ dipole
propagating in a medium might be difficult to understand, a more intuitive
derivation based on the classic multiple scattering theory of Moli\`ere is
presented in Section~\ref{intuition}. It goes along with the treatment of
multiple interactions in the abelian case by Levin and Ryskin \cite{lr}.
Although it does not have explicit QCD input, the final result has the form
of a color-screened dipole cross section, and one can see where different
parts of this cross section come from. One can also trace the origin of the
$A^{1/3}$-dependence of $\la k_T^2\ra$ broadening to multiple interactions
and see that the lowest-order contribution actually has a quite different
behavior.

The energy-dependent dipole cross section is evaluated in
Section~\ref{c} via the unintegrated gluon density, which
is estimated within models adjusted to describe data
for the proton structure function over a wide range of $x$ and $Q^2$,
as well as for the hadronic total cross sections.
The predicted broadening is expected to rise steeply with energy.

Since $k_T$ broadening of a high energy parton results from multiple
gluonic exchanges with nucleons, nuclear shadowing of 
gluons diminishes the effect of broadening. This can be also interpreted as
the Landau-Pomeranchuk effect, or coherence-suppressing
gluon radiation, which gives an important contribution to the broadening.
In Section~\ref{gluons} we evaluate the effect of shadowing 
relying on the light-cone Green function
formalism, which includes the strong nonperturbative interaction
of gluons that dramatically diminishes the effect of shadowing.
The reduction of broadening due to gluon shadowing turns
out to be rather small, less than $10\%$.

Nuclear broadening of $k_T$ can be affected also by the finiteness
of the experimental aperture. In Section~\ref{aperture} we estimate the 
corresponding correction factor, which is rather close to unity
if the aperture covers a range of $k_T$ exceeding a few $GeV$.

The main observations are summarized and discussed in the last section.

\section{Transverse momentum broadening as color filtering}\label{filter}

Nuclear broadening of the transverse momentum of a quark
propagating through a medium was treated in \cite{dhk} in terms of eikonalized
multiple interactions of a colorless $\bar qq$ pair via gluon exchanges. 
It was found that the mean transverse 
momentum squared grows as
\beq
\delta\la k_T^2\ra=
2\,C\,\rho_A\,L\ .
\label{3}
\eeq
Here $L$ is the length of the path covered by
the quark in nuclear matter up to the point 
that the lepton pair
is produced, and  $\rho_A$ is the nuclear density. 
The factor $C$ originates from the expression for the 
total cross section for the interaction between a nucleon and 
a colorless $\bar qq$ dipole having transverse 
separation $r_T$ and c.m. energy squared $s$ \cite{zkl},
\beq
\sigma_{\bar qq}(r_T,s) = C(r_,s)\,r_T^2\ .
\label{4}
\eeq
The $r_T^2$ behavior at small $r_T\to 0$ is dictated by
gauge invariance and the non-abelian nature of QCD. The energy-independent 
part of $C$ was estimated
in the Born approximation to be $C\approx 3$ \cite{zkl,kz}.
However higher order perturbative corrections lead to a rising
energy dependence of the $C(r_T,s)$.
At small $r_T$ the factor $C(r_T,s)$ is related to the gluon density
in the proton \cite{fs,nz94},
\beq
C(r_T,s)=\frac{\pi^2}{3}\,G(x,Q^2)\ ,
\label{4a}
\eeq
where $G(x,Q^2)=x\,g(x,Q^2)$ is the gluon distribution function
which depends on $Q^2\sim 1/r_T^2$ and $x=Q^2/s$.  The
the broadening corresponding to (\ref{3}) takes the form,
\beq
\delta\la k_T^2\ra = \frac{2\,\pi^2}{3}\,G(x,Q^2)\,\rho_A\,L\ .
\label{5}
\eeq
This expression is also derived in \cite{baier}.

Although the result (\ref{3}) is correct it was poorly motivated in
\cite{dhk}. Better derivations one can find in \cite{baier,wg}. They,
however, assume Gaussian shape for the $k_T$ distribution and employ the
approximation of constant $C(r_T)$ which cannot be justified for soft
multiple interactions. Here we derive a general expression (\ref{a.12}) for
nuclear modification of the transverse momentum distribution which is free
of these approximations.

The effect of the mean-square transverse momentum of the quark
after propagation through a medium depends on the
reaction, nevertheless, the broadening $\delta\la k_T^2\ra$ 
is universal. This is demonstrated below for the example 
of a hadron-nucleus interaction where we are
interested in the final transverse momentum distribution
of one of the projectile quarks. This can be expressed in terms
of the density matrix of the final quark,
$\Omega^q_f(\vec b,\vec b')$, where $\vec b$ is an impact parameter,
\beq
\frac{dN_q}{d^2k_T} =
\int d^2b\,d^2 b'\,{\rm exp}\,\Bigl[
i\,\vec k_T\,(\vec b-\vec b')\Bigr]\,
\Omega^q_f(\vec b,\vec b')\ .
\label{a.1}
\eeq
The distribution $dN_q/d^2k_T$ is normalized to one.
The final density matrix is related to the initial one
as
\beq
\Omega^q_f(\vec b,\vec b')={\rm Tr}\,
\widehat S^{\dagger}(\vec b'+\vec B)\,
\hat \Omega^q_{in}(\vec b,\vec b')\, \widehat S(\vec b+\vec B)\ .
\label{a.2}
\eeq
Here $\widehat S(\vec b+\vec B)$ is the $S$-matrix for a quark-nucleus
collision with impact parameter $\vec b+\vec B$ where $\vec B$
and $\vec b$ are the impact parameters between the center of
gravity of the projectile hadron and the center of the nucleus
or the quark, respectively. We take the trace
over the color indices of the quark.
The initial density matrix reads,
\beq
\hat \Omega^q_{in}(\vec b_1,\vec b_1') =
\sum\limits_{n,\,\{in\}} |C_n|^2\,
\int d^2b_2\,d^2b_3\,...\,d^2b_n\,
\Psi^{\dagger}_n(\vec b_1,\vec b_2,...,\vec b_n)\,
\Psi_n(\vec b_1',\vec b_2,...,\vec b_n)\ .
\label{a.3}
\eeq
Here we sum over different Fock components of the hadron
containing different numbers $n$ of (anti)quarks with
weight factors $|C_n|^2$. We also sum over the initial
state polarizations and colors of all quarks except for
the first one. This makes the matrix $\Omega^q_{in}(\vec b_1,\vec b_1')$
diagonal in color space. We do not show explicitly the
dependence of the hadron light-cone wave
function on the longitudinal momenta of the quarks, assuming
integration over all longitudinal momenta except for that
of the first quark. In the high energy approximation we neglect
the energy loss of the quark propagating over a finite
path in the nuclear medium. This effect, if it becomes
important, should be treated separately.

Let us consider the $S$-matrix of a quark-nucleus
collision in the approximation where all the coordinates
$\vec r_i$ of the bound nucleons, as well as the intrinsic
quark coordinates $\vec\rho_j$ in the nucleons, are ``frozen'' during
the interaction time.
In this case the $S$ matrix acquires the eikonal form,
\beqn
\widehat S(\vec b+\vec B,\vec r_i;\{\vec\rho,\mu\}_i) &=&
\sum\limits_P
\Theta(z_2-z_1)\,...\,\Theta(z_A-z_{A-1})\nonumber\\
&\times&\hat s_1(\vec b+\vec B -\vec r_{T_1};\{\vec\rho,\mu\}_1)\,...\,
\hat s_A(\vec b+\vec B -\vec r_{T_A};\{\vec\rho,\mu\}_A)\ .
\label{a.4}
\eeqn
Here we sum over permutations of the nucleons.
$\{\vec\rho,\mu\}_i$ denotes the set of intrinsic quark coordinates
and color indices in the $i$-th nucleon.
The single quark-nucleon $S$-matrix reads,
\beq
\hat s_i(\vec b+\vec B-\vec r_{T_i};\{\vec\rho,\mu\}_i)=
{\rm exp}\,\left[\frac{i}{4}\,
\sum\limits_{j=1}^3\hat\lambda_a\,
\hat\lambda_a(j)\,\chi(\vec b+\vec B -\vec r_{T_j})\right]\ ,
\label{a.5}
\eeq
where $\hat\lambda_a$ are the Gell-Mann matrices, and index
$j$ refers to one of the quarks in the target nucleon;
\beq
\chi(\vec b)= \int\limits_{\Lambda^2}^{\infty}
\frac{dq^2}{q^2}\,\alpha_s(q^2)\,J_0(b\cdot q)\ .
\label{a.6}
\eeq
Here $J_0$ is a Bessel function, $\vec q$ is the transverse momentum
of the gluon exchanged in the $t$-channel, and $\Lambda^2$ is an infra-red
cut off.

\begin{figure}[tbh]
\includegraphics{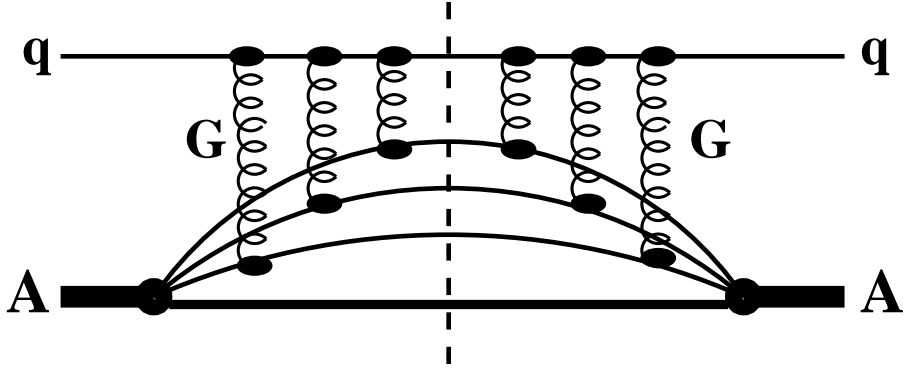}
\begin{center}
\vspace{5cm}
\parbox{13cm}
{\caption[Delta]
{\sl The probability of multiple interactions
via one gluon exchange for the
quark in the nucleus. The dashed line shows the unitarity 
cut.}
\label{fig1}}
\end{center}
\end{figure}

As soon as the initial density matrix $\hat \Omega^q_{in}(\vec b,\vec b')$
is diagonal in the color indices of the quark (see above),
we can average the product $\widehat S^{\dagger}\widehat S$
over the colors and coordinates of the quarks in the target
nucleons in inverse sequence starting from the last one.
Then we average over the positions of the nucleons $\{\vec r\}$,
\beqn
&\biggl\la\left\la \widehat                   
S^{\dagger}(\vec b'+\vec B,\vec r_i;\{\vec\rho,\mu\}_i)\,
\widehat S(\vec b+\vec B,\vec r_i;
\{\vec\rho,\mu\}_i)\right\ra_{\{\vec\rho,\mu\}}
\biggr\ra_{\{\vec r\}}
= \Biggl\la\sum\limits_P
\Theta(z_2-z_1)\,.\,.\,.\,\Theta(z_A-z_{A-1})&
\nonumber\\
&\times\left\la
\hat s_1^{\dagger}(\vec b'+\vec B-\vec r_{T_1})
\left\la\hat s_2^{\dagger}(\vec b'+\vec B-\vec r_{T_2})\,.\,.\,.\,
\left\la\hat s_A^{\dagger}(\vec b'+\vec B-\vec r_{T_A})
\cdot\hat s_A(\vec b+\vec B-\vec r_{T_A})
\right\ra_{\{\vec\rho,\mu\}_A}\,.\,.\,.\,
\right.\right.\nonumber\\
&\times\left.\left.
\hat s_2(\vec b+\vec B-\vec r_{T_2})\right\ra_{\{\vec\rho,\mu\}_2}
\hat s_1(\vec b+\vec B-\vec r_{T_1})\right\ra_{\{\vec\rho,\mu\}_1}
\Biggr\ra_{\{\vec r\}}\ .
\label{a.7}
\eeqn
This expression is illustrated in Fig.~\ref{fig1}, where we use
the two-gluon model for the Pomeron for the sake of simplicity.
One can see that the Pomerons are enclosed within each other.

After averaging over the target nucleon wave function including
the target quark color indices, the functions
\beq
U_k(\vec b'+\vec B-\vec r_{T_k},\vec b+\vec B-\vec r_{T_k})=
\left\la\hat s^{\dagger}_k(\vec b'+\vec B-\vec r_{T_k})
\cdot\hat s_k(\vec b+\vec B-\vec r_{T_k})
\right\ra_{\{\vec\rho,\mu\}_k}\ 
\label{a.8}
\eeq
commute, and the sum over permutations of the nucleon coordinates
of the product of $\Theta$-functions equals one.
This expression becomes independent of index $k$ after integration over
$\vec r_k$. The integral can be represented as,
\beqn
&&\frac{1}{A}\int d^3r\,\rho_A(\vec r_T,z)\,
U(\vec b'+\vec B-\vec r_{T},\vec b+\vec B-\vec r_{T})
\nonumber\\&=&
\frac{1}{A}\int d^3r\,\rho_A(\vec r_T,z) -
\frac{1}{A}\int d^3r\,\rho_A(\vec r_T,z)\,
\Bigl[1-U(\vec b'+\vec B-\vec r_{T},
\vec b+\vec B-\vec r_{T})\Bigr]
\nonumber\\&\approx&
1 - \frac{1}{2\,A}\,
T_A\left(\frac{\vec b+\vec b'}{2}+\vec B\right)\,
\sigma_{\bar qq}(\vec b-\vec b')\ .
\label{a.9}
\eeqn
We use the standard approximation of uncorrelated
nuclear wave functions, and $\rho_A(\vec r)$ is the
one-body nuclear density normalized to $A$.
$T_A(\vec b)$ is the nuclear thickness function,
\beq
T_A(b)=\int\limits_{-\infty}^{\infty} dz\,\rho_A(b,z)\ ,
\label{6}
\eeq
and 
\beq
\sigma_{\bar qq}(\vec\rho)=
\frac{2}{A}\int d^2r_T\,
\Bigl[1-U(\vec b'+\vec B-\vec r_{T},\vec b+\vec B-\vec r_{T})\Bigr]
\label{a.11}
\eeq
is the total cross section for the interaction of a $\bar qq$ dipole of
transverse separation $\vec\rho=\vec b-\vec b'$ \cite{zkl} with a nucleon.

Eventually we arrive at the expression for the transverse
momentum distribution of a quark that has propagated through the nucleus,
\beq
\frac{dN_q}{d^2k_T} =
\int d^2b\,d^2 b'\,{\rm exp}\,\Bigl[
i\,\vec k_T\,(\vec b - \vec b')\Bigr]\,
\Omega^q_{in}(\vec b,\vec b')\,
{\rm exp}\,\left[-{1\over2}\,
\sigma_{\bar qq}(\vec b-\vec b')\,
T_A\left(\frac{\vec b+\vec b'}{2}+\vec B\right)
\right]\ .
\label{a.12}
\eeq
This equation describes the full $k_T$-distribution of the
final quarks, as well as 
the mean transverse momentum squared of an ejectile  
quark in a hadron-nucleus collision, 
\beqn
\la k_T^2\ra_{hA} &=& 
\frac{1}{A}\int d^2B\,\int\limits_{-\infty}^{\infty} dz\,
\rho_A(B,z)\int d^2k_T\,k^2_T\int d^2b_1\,d^2b_2\,
\Omega^q_{in}(\vec b_1,\vec b_2)
\nonumber\\&\times& {\rm exp}\,
\left[i\,\vec k\,(\vec b_1-\vec b_2)\right]\,
{\rm exp}\,\left[-{1\over 2}
\sigma_{\bar qq}(\vec b_1-\vec b_2)\,
T_A\left(\frac{\vec b_1+\vec b_2}{2}+\vec B,z \right)\right]\ .
\label{5a}
\eeqn

The integration in (\ref{5a}) is easy to perform by replacing
$k_T^2\,{\rm exp}\,[i\,\vec k\,(\vec b_1-\vec b_2)]$ by
derivatives  
$\vec\nabla(b_1)\cdot\vec\nabla(b_2)\,
{\rm exp}\,[i\,\vec k\,(\vec b_1-\vec b_2)]$. 
Integrating by parts we arrive at the expression
\beqn
&&\la k_T^2\ra_{h A} =
\frac{1}{A}\int d^2B\,\int\limits_{-\infty}^{\infty} dz\,
\rho_A(B,z)\int d^2b_1\,
\left\{\vec\nabla(b_1)\cdot\vec\nabla(b_2)\,
\Omega^q_{in}(\vec b_1,\vec b_2)\right.\nonumber\\
&+& \left.
\vec\nabla(b_1)\cdot\vec\nabla(b_2)\,
{\rm exp}\,\left[-{1\over 2}\,
\sigma_{\bar qq}(\vec b_1-\vec b_2)\,
T_A(B,z)\right]\right\}_{\vec b_2=\vec b_1}\ ,
\label{7}
\eeqn
where we assume that the nuclear radius is much larger 
than that of the the projectile hadron.

Using a Gaussian for the hadronic wave function,
the density matrix of the initial quark in the case of a proton
beam reads,
\beq
\Omega^q_{in}(\vec b_1,\vec b_2) =
\frac{2}{3\pi\la r^2_{ch}\ra}\,
{\rm exp}\,\left[- \frac{b_1^2+b_2^2}
{3\,\la r^2_{ch}\ra}\right]\ ,
\label{8}
\eeq
where $\la r^2_{ch}\ra$ is the mean square of the proton charge 
radius. Note that we do not introduce any unreasonably large
primordial transverse momentum for the projectile quark, which
is usually assumed for hard reactions in leading order
in the parton model. In the light cone approach,
the higher order perturbative corrections are already included
in the phenomenological cross section (\ref{4}), and they generate
the energy dependence of $C(r_T,s)$. 

In the second term on the {\sl r.h.s.} of (\ref{7}) one should 
apply both derivatives to $\sigma_{\bar qq}(\vec b_1-\vec b_2)$
in the exponential. This is because $\sigma_{\bar qq}(\vec b_1-\vec b_2)
\propto (\vec b_1-\vec b_2)^2$ as $\vec b_1 \to \vec b_2$,
otherwise the result is zero.
Using Eq.~(\ref{4}) for the dipole cross section, one gets
from (\ref{5a}) the mean-square transverse momentum of
a valence quark,
\beq
\la k_T^2\ra_{pA} = \frac{2}{3\la r^2_{ch}\ra} +
2\,C(0,s)\,\la T_A\ra\ ,
\label{9}
\eeq
where 
\beq
\la T_A\ra = \frac{1}{A}\int d^2b\,T^2(b)
\label{10}
\eeq
is the mean nuclear thickness.
Thus, with a better derivation we confirm the result of \cite{dhk}, 
Eq.~(\ref{3}).

\section{Moli\`ere theory: a more intuitive derivation}\label{intuition}

The result we obtained in Eqs.~(\ref{7}), (\ref{9}) might look puzzling.
Indeed, in the expansion of the exponential in (\ref{7}) only the lowest
order term $\propto \sigma_{\bar qq}\,T_A\, \propto\, A^{1/3}$ survives
differentiation and contributes to $\la k_T^2\ra$. This observation seems
to lead to the conclusion that only single rescattering contributes to $\la
k_T^2\ra$, while higher multi-fold scatterings do not affect the broadening
of the transverse momentum (compare with \cite{qs}).  Such an
interpretation contradicts the intuitive expectation that broadening is
caused by multiple interactions in the medium resulting in random walk of
the particle in the transverse momentum plane. Here we present a different,
more intuitive derivation for the same result (\ref{9}) based on
Moli\`ere's theory \cite{moliere} of multiple interactions in a medium (see
also \cite{lr}).  This result shows that broadening is indeed caused by
multiple interactions rather than by the single rescattering term, which
has an $A$-dependence very different from $A^{1/3}$.

We concentrate on nuclear broadening of $\la k_T^2\ra$ and for the sake of
clarity neglect the primordial transverse momentum distribution of the
projectile quark which is responsible for the first term on the {\it
r.h.s.} of (\ref{9}). Evolution of the transverse momentum distribution
$D(k_T,z)$ as function of longitudinal coordinate $z$ is described by the
kinetic equation,
 \beq
\frac{d\,D(k_T,z)}{d\,z} = 
-\,\sigma_{tot}\,\rho_A(z)\,D(k_T,z) 
+ \int d^2 k_T^\prime\,
\rho_A(z)\,\frac{d\,\sigma(\vec k_T - \vec k_T^\prime)}
{d^2k_T^\prime}\,D(k_T^\prime,z)\ .
\label{20}
\eeq
Here $d\sigma/d^2k_T$ is the differential cross section of
quark scattering on a nucleon summed over final states and
quark colors. The first term on the {\it r.h.s.} of
(\ref{20}) and the total cross section,
\beq
\sigma_{tot}=\int d^2 k_T\,
\frac{d\,\sigma}{d^2k_T}\ ,
\label{21}
\eeq
describe the leakage of quarks from the element of
phase space $d^2k_T$. At the same time, this region 
gains new quarks from other parts of phase space  
via scattering as is described by the second term in (\ref{20}).

Equation~(\ref{20}) is easy to solve switching to coordinate space,
\beq
D(k_T,z)=\frac{1}{2\,\pi}
\int d^2r_T\,e^{i\,\vec r_T\,\vec k_T}\,
D(r_T,z)\ .
\label{22}
\eeq
The solution is (up to the initial condition),
\beq
D(k_T,z) \propto \frac{1}{2\,\pi}
\int d^2r_T\,e^{i\,\vec r_T\,\vec k_T}\,
\exp\left[-\Bigl(\sigma_{tot} - \gamma(r_T)\Bigr)
\,T_A(z)\right]\ ,
\label{23}
\eeq
where
\beq
\gamma(r_T) = \int d^2k_T\,e^{-i\,\vec k_T\,\vec r_T}\,
\frac{d\,\sigma}{d\,k_T^2}\ ,
\label{24}
\eeq
\beq
\gamma(0)=\sigma_{tot}\ ,
\label{24a}
\eeq
\beq
T_A(z)=\int\limits_{-\infty}^{z} dz'\,\rho_A(z')\ .
\label{24b}
\eeq

We are now in a position to discuss the meaning of expansion
of the exponential in (\ref{23}). First of all, the expansion,
\beq
\exp(-\sigma_{tot}\,T_A) = 
\sum\limits_{n=0} \frac{(-1)^n}{n!}\,
(\sigma_{tot}\,T_A)^n\ ,
\label{25}
\eeq
cannot be treated as a multiple scattering series.
This is just an attenuation exponential, and the particle
(quark) cannot be absorbed (knocked out of a phase space cell)
twice. This simple exponential should not be confused
with the Glauber eikonal multiple scattering series, which
has a nontrivial interpretation in terms of unitarity relation
and AGK cutting rules \cite{agk}.

At the same time, the expansion of the second term, $\gamma(r_T)\,T_A$,
in the exponent of (\ref{23}) does have the interpretation of a 
multiple scattering series. Indeed,
\beqn
&&\int d^2r_T\,\exp(i\,\vec k_T\cdot\vec r_T) 
\exp\Bigl[\gamma(r_T)\,T_A\Bigr] 
\nonumber\\
&=&\, \delta(\vec k_T) 
\,+\, \sum\limits_{n=1}\,
\frac{T_A^n}{n!}\,
\int \prod\limits_{j=1}^nd^2k_j\,
\frac{d\sigma(\vec k_j)}{d^2k_j}\,\,
\delta\left(\vec k_T-\sum\limits_{i=1}^n
\vec k_i\right)\ .
\label{26}
\eeqn
The $n$-th term on the sum in the {\it r.h.s.} of
this equation clearly corresponds to the $n$-fold scattering of the
quark. Thus, all multi-fold rescatterings contribute to the
shape of the $k_T$-distribution of a quark propagating through a
medium. 

Amazingly, the exponent in (\ref{23}) can be represented as a color dipole
cross section, making it similar to the results of previous section,
\beq
\sigma_{tot} - \gamma(r_T) = 
\frac{1}{2\,\pi} \int d^2k_T\,
\left(1 - e^{i\,\vec k_T\cdot\vec r_T}\right)\,
\frac{d\,\sigma}{d^2k_T} = {1\over2}\,
\sigma_{\bar qq}(r_T)\ .
\label{27}
\eeq
We have made use of the fact that $d\sigma/d^2k_T\propto 1/k_T^4$.
This expression is also derived in more detail at the end of
this section.

The appearance of the dipole cross section, as in (\ref{5a}), is
the result of an artificial construction.  Clearly, the object participating
in the scattering is not the colored $\bar qq$ dipole but rather a single 
colored quark.  Now it turns out that the different parts of this 
dipole cross section have quite different origins.  Namely,
the first term in the brackets in (\ref{27}) (the ``1") corresponds
to simple attenuation of the projectile quark detected in
a given phase space cell. However, the second term
($\exp(i\,\vec k_T\cdot\vec k_T)$) originates from multiple scattering
of the quark.

Thus, if one needs to establish a relation between 
the expansion of the exponential
$\exp[-{1\over2}\sigma_{\bar qq}(r_T)\,T_A]$ in (\ref{5a}) 
and the multiple quark
interaction, it would be incorrect to think that the $n$-th order term of
this expansion corresponds to the probability $W_n$ to have $n$-fold
quark multiple scattering,
\beq
W_n(r_T)\ \not=\ \frac{(-1)^n}{n!}\,
\Bigl[\sigma_{\bar qq}(r_T)\,T_A\Bigr]^n\ ,
\label{28}
\eeq
but, instead,
\beq
W_n(r_T) = e^{-\sigma_{tot}\,T_A}\,
\frac{\gamma^n(r_T)T_A^n}{n!}\ .
\label{29}
\eeq
In contrast to (\ref{28}), all the terms in (\ref{29}) 
are positive as they should be. This standard result of the multiple
scattering theory has been also revived recently in \cite{urs}.

Now we are in position to figure out where the mean $k_T^2$ comes from.
We find
\beq
\la k_T^2\ra = \frac{1}{\sigma_{tot}}
\int d^2k_T\,k_T^2 \int d^2r_T\,
e^{i\,\vec k_T\cdot\vec r_T}\, 
\sum\limits_{n=0}W_n(r_T)\ =\ 
k_0^2\,\la n\ra\ ,
\label{30}
\eeq
where
\beq
k_0^2 = \frac{1}{\sigma_{tot}}
\int d^2k_T\,k_T^2\,
\frac{d\,\sigma}{d\,k_T^2}\ ,
\label{31}
\eeq
is the mean transverse momentum squared gained by the quark in a single
scattering, and
\beq
\la n\ra = \sigma_{tot}\,T_A
\label{32}
\eeq
is the mean number of interactions of the quark
propagating though nuclear thickness $T_A$.

Equation~(\ref{30}) explicitly demonstrates that nuclear broadening
of the quark mean transverse momentum is the result of multiple interactions
leading to a random walk in the transverse momentum plane. Therefore, it would
be incorrect to interpret the result of the previous section,
Eq.~(\ref{5a}), as the contribution of a single scattering, which 
must contain an extra factor
$\exp(-\sigma_{tot}\,T_A)$ and has quite a different
$A$-dependence compared to $T_A\propto A^{1/3}$ of Eq.~(\ref{5a}).

Concluding this section we present a derivation of Eqs.~(\ref{5a}) - 
(\ref{9}) within the model of potential scattering.
The amplitude for elastic scattering of a particle from a potential
reads,
\beq
f(k_T)=\frac{i}{2\,\pi}
\int d^2b\,e^{i\,\vec b\cdot\vec k_T}\,
\omega(b)\ ,
\label{33}
\eeq
where
\beq
\omega(b)=1-e^{i\,\chi(b)}\ ;
\label{34}
\eeq
\beq
\chi(b) = - {1\over v}\int\limits_{-\infty}^{\infty}
dz\,V(\vec b,z)\ .
\label{35}
\eeq
Correspondingly,
\beq
\sigma_{tot}=4\,\pi\,{\rm Im} f(0) = 
2\,{\rm Re} \int d^2b\,\omega(b)\ ;
\label{36}
\eeq
\beq
\frac{d\sigma}{d^2k_T} =
\Bigl|f(k_T)\Bigr|^2\ ;
\label{37}
\eeq
\beq
\sigma_{el} = \int d^2k_T\,
\frac{d\sigma}{d^2k_T} = 
\int d^2b\,\Bigl(2-2\,\cos \chi(b)\Bigr)
= \sigma_{tot}\ ,
\label{38}
\eeq
as one should have expected for potential scattering.

Substituting (\ref{37}) into (\ref{24}) we get,
\beq
\gamma(r_T)=\int d^2b\,\omega^*\Bigl(\vec b+\alpha\vec r_T\Bigr)\,
\omega\Bigl(\vec b-(1-\alpha)\vec r_T\Bigr)
=\sigma_{tot} - {1\over2}\,\sigma_{\bar qq}(r_T)\ ,
\label{39}
\eeq
where $0<\alpha <1$ is an arbitrary number, and 
\beq
\sigma_{\bar qq}(r_T)=
2\,{\rm Re}\int d^2b\,\left\{1-
\exp\left[i\,\chi\Bigl(\vec b+\alpha\vec r_T\Bigr)
-i\,\chi\Bigl(\vec b-(1-\alpha)\vec r_T\Bigr)\right]\right\}
\label{40}
\eeq
is the total cross section for a quark-antiquark pair with transverse
separation $r_T$, called dipole cross section. In this case $\alpha$
can be interpreted as a share of the total light-cone momentum carried 
by a quark or antiquark.
Thus, we confirm  Eq.~(\ref{27}).

\section{Calculation of the \boldmath$C(0,s)$}\label{c}

As mentioned above, all total cross sections rise with
energy, so the factor $C(r_T,s)$ does also. The energy dependence
is steeper towards small $r_T$ \cite{kp1} and may lead to dramatic 
changes in $C(r_T,s)$ compared to the oversimplified estimate
$C \approx 3$ \cite{zkl,kz}. 

Intuitively, it is clear that a fast quark scatters off the
gluon clouds of bound nucleons. The gluon density at small 
Bjorken $x$ relevant to such a high-energy interaction increases
with $1/x$, {\it i.e.} with quark energy (see Eq.~\ref{5}). 
Said differently, multiple interactions of the quark
are accompanied by gluon bremsstrahlung.  As the energy is increased,
more phase space becomes available for the radiated gluons.
Since gluon radiation enhances transverse motion of the
quark, $\la k_T^2\ra$ should grow with energy (via $C(0,s)$).

Thus,
the energy dependence of $C(r_T,s)$ can be expressed in terms
of the unintegrated gluon density \cite{nz94},
\beq
C(0,s) = \frac{\pi}{3}\int
d^2k\,\frac{\alpha_s(k^2)\,k^2}{k^4}\,
{\cal F}(x,k^2)\ .
\label{11}
\eeq
where ${\cal F}(x,k^2)=
\partial G(x,k^2)/\partial({\rm ln}k^2)$ and
$G(x,Q^2)=x\,g(x,Q^2)$.
Here $\vec k$ is the transverse momentum that a quark acquires 
scattering off the gluon cloud of a nucleon. $\alpha_s(k^2)$
is the QCD running constant, which we calculate in  the one-loop 
approximation. 

The value of $x$ for gluons in the {\it r.h.s.} of (\ref{11})
is related to the quark-nucleon c.m. energy squared $s$.  
The minimal value of $x$ corresponding to a collinear quark-gluon 
collision reads,
\beq
x_{min}=\frac{4\,k^2}{s}\ .
\label{x-s}
\eeq 

The density of the gluon cloud around the proton
should vanish at large distances because of the color neutrality
of the proton. Therefore, the unintegrated gluon distribution
${\cal F}(x,k^2) \propto k^2$ as $k^2\to 0$. This provides
infra-red stability of the integral in (\ref{11}).

There are quite a few models for the unintegrated gluon 
distribution ${\cal F}(x,k^2)$ at high $k^2$ 
({\it e.g.} see \cite{kms});
however, very little information is available in 
region of small $k^2$ where perturbative QCD cannot be
used. Since the integral in
(\ref{11}) is dominated by soft gluons, one should
develop a phenomenology for ${\cal F}(x,k^2)$
and use other observables
to restrict possible uncertainties.
One of the sensitive probes for ${\cal F}(x,k^2)$
(even more sensitive than $C(0,s)$) is the total hadronic cross
section. The simplest case is 
the pion-proton cross section, which is given
by the same approach as Eq.~(\ref{11}),
\beq
\sigma^{\pi p}(s) = \frac{4\,\pi}{3}\int
d^2k\,\frac{\alpha_s(k^2)}{k^4}\,
\Bigl[1-F^{\pi}_{qq}(k)\Bigr]\,{\cal F}(x,k^2)\ .
\label{13}
\eeq
Here $F^{\pi}_{qq}(k)=\left\la \pi\left|{\rm exp}
\Bigl[i\,\vec k\,(\vec r_2-\vec r_1)
\Bigr]\right|\pi\right\ra \approx 
{\rm exp}(-k^2\,\la r_{\pi}^2\ra/3)$ 
is the two-quark
formfactor of the pion, and $\la r_{ch}^2\ra_{\pi} =
0.44\pm 0.01\,fm^2$
\cite{pion} is the mean-square pion charge radius.

\begin{figure}[tbh]
\includegraphics{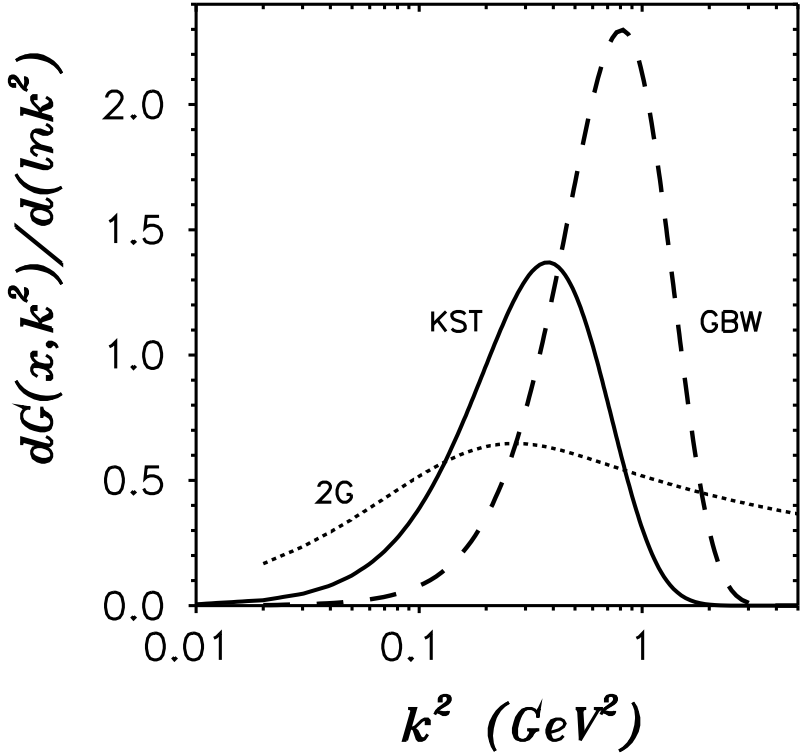}
\begin{center}
\vspace{9.5cm}
\parbox{13cm}
{\caption[shad1]
{\sl The non-integrated gluon density 
${\cal F}(x,k^2)$ calculated at the quark-nucleon energy $s=500\,GeV^2$
as function of $k^2$. 
The dashed and solid curves
correspond to the models GBW and KST.
The dotted curve shows the two-gluon contribution
(\ref{12}).}
\label{gg}}
\end{center}
\end{figure}

In the Born approximation \cite{l,n,zkl} the gluon density 
takes a simple form,
\beq
{\cal F}(x,k^2) \Rightarrow
{4\over\pi}\,\alpha_s(k^2)\,
\Bigl[1-F^N_{qq}(k)\Bigr]\ ,
\label{12}
\eeq
where $F^p_{qq}(k)=\left\la N\left|{\rm exp}
\Bigl[i\,\vec k\,(\vec r_2-\vec r_1)
\Bigr]\right|N\right\ra \approx {\rm exp}(-k^2\,\la r_{ch}^2\ra_p/2)$ 
is the two-quark 
formfactor of the nucleon, and $\la r_{ch}^2\ra_p = 0.79\pm 0.03\,fm^2$
\cite{garching} is the mean-square charge radius of the proton.
In fact, because Coulomb gluons have no partonic interpretation, quark 
elastic scattering through the exchange of two such gluons cannot be 
expressed in terms of the gluon density as given in Eq.~(\ref{5}).
 Therefore, 
(\ref{12}) cannot be treated as a model for the gluon distribution function 
${\cal F}(x,k^2)$ by any means; nevertheless we plot it {\it vs} $k^2$ in
Fig.~\ref{gg} (as the dotted curve) to compare with other models.
This contribution is independent of $x$. As was mentioned, $x$-dependence
originates from  gluon bremsstrahlung.

Comparing the factor $(1-F^{\pi}_{qq}(k))$ in (\ref{13}) to
$k^2$ in (\ref{11}), it is clear that the total hadronic cross section
is much more sensitive to the behavior of
${\cal F}(x,k^2)$ than $C(0,s)$ is
in the region of interest, namely at small $k^2$.
Therefore, reproduction of the total 
hadronic cross section is an important test for
any model of the gluon density. At large values $k^2$ perturbative QCD
is supposed to be valid, and different models should therefore converge
there.

Note that one should interpret (\ref{13}) 
as the inelastic rather than the total cross section
because it is linear in gluon density, {\it i.e.}
corresponds to a color octet-octet unitarity cut, which
does not contain elastic or diffractive states.
Therefore, one should add to (\ref{13}) the elastic and single 
diffraction cross sections to get the total cross section. However,
such an amplitude linear in the gluon density is subject to
unitarization. The first unitarity correction, which is
quadratic in the gluon density, is just the same elastic and 
diffractive cross sections, but with a negative sign.
Hence, this cancels the contributions of the elastic and diffractive
cross sections, and one can treat (\ref{13}) as the total
cross section
neglecting the higher order unitarity corrections $O(G^3)$,
which are known to be quite small at the present energies.

We will try a few models for the gluon density
to calculate the $C(0,s)$, hoping that the spread in
the results may serve as a measure for the theoretical uncertainty.

\subsection{GBW model}\label{gbw}

Golec-Biernat and W\"usthoff \cite{gw} suggested a model
for the dipole cross section (\ref{4}) which
saturates at large $\bar qq$
separations and reproduces well the DIS cross sections.
\beq
\sigma_{\bar qq}(r_T,x)=\sigma_0\,\left[
1-{\rm exp}\left(-\frac{r_T^2}
{R_0^2(x)}\right)\right]\ ,
\label{16}
\eeq
where $R_0(x)=0.4\,fm \times (x/x_0)^{\lambda/2}$ and
$\sigma_0=23.03\,mb$; $\lambda=0.288$; $x_0=3.04\cdot 10^{-4}$.
This cross section obviously satisfies the behavior
suggested by data for $F_2(x,Q^2)$: the smaller the separation $r_T$, 
the steeper the growth with $1/x$. However, the value of $x$ is
not well defined for a dipole of a given energy (there is no problem
with the definition of $x$ in our next model).

The gluon density corresponding to this dipole cross section reads
\cite{gw},
\beq
{\cal F}(x,k^2) = 
\frac{3\,\sigma_0}{16\,\pi^2\,\alpha_s(k^2)}\
k^4\,R_0^2(x)\,{\rm exp}\Bigl[-{1\over4}\,R_0^2(x)\,k^2\Bigr]\ .
\label{17}
\eeq
This function is depicted in Fig.~\ref{gg} by the dashed curve.
Note that according to Eq.~(\ref{17}) 
the right wing of this curve at large $k^2$ is a steeply
rising function of energy.

This model faces obvious problems with hadronic cross sections.   
Indeed, the dipole cross section (\ref{16}) grows steeply                 
with $r_T$ and saturates for $r_T > R_0(x)$ at    
$\sigma_{\bar qq}(r_T,x) \leq \sigma_0$. Averaging (\ref{16}) weighted
with the pion wave function squared, one never can reach a pion-proton
total cross section larger than $\sigma_0=23.03\,mb$.
%
%Indeed, the predicted total cross section $\sigma^{\pi p}=
%20.2\,mb\,(s/1000\,GeV^2)$ is nearly constant at higher energies.
               
\begin{figure}[tbh]
\includegraphics{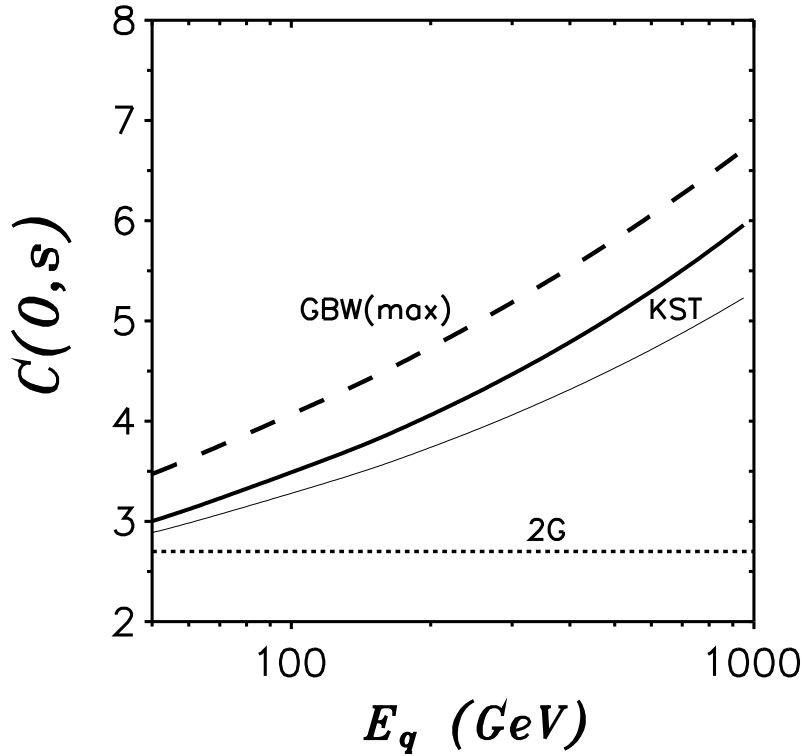}
\begin{center}
\vspace{9.5cm}
\parbox{13cm}
{\caption[shad1]
{\sl Factor $C(0,s)$ as function of the quark 
energy calculated with (\ref{11}). The curve assignments 
is the same as in Fig.~\ref{gg}, and the thin solid curve
is the KST curve corrected for gluon shadowing in lead.}
\label{c(0)}}
\end{center}
\end{figure}

In order to calculate the energy dependence of 
$C(0,s)$ we
need to know the value of $x$, which is poorly defined
in this model. If we use the minimal value (\ref{x-s}) permitted by
kinematics, the result shown by the dashed curve in Fig.~\ref{c(0)}
should be an upper bound for $C(0,s)$ in this model.
We see that $C(0,s)$ is rather steep as a function of energy.

\subsection{KST model}\label{kst}

The advantage of the GBW parameterization (\ref{16}) is simplicity and
convenience for analytical calculations. One may keep this form,
but modify it as suggested in \cite{kst} to address the problems mentioned 
above.  An explicit energy dependence is introduced in the
parameter $\sigma_0$ in (\ref{16}) in a way that guarantees the
reproduction of the correct hadronic cross sections,
\beq
\sigma_0(s)=\sigma^{\pi p}_{tot}(s)\,
\left(1 + \frac{3\,R^2_0(s)}{8\,\la r^2_{ch}\ra_{\pi}},
\right)
\label{18}
\eeq
where $\sigma^{\pi p}_{tot}(s)=23.6\times(s/s_0)^{0.08}$ 
is the Pomeron part
of the $\pi p$ total cross section \cite{pdt}, and
$R_0(s)= 0.88\,fm \times(s/s_0)^{-\lambda/2}$ with $\lambda=0.28$ 
and $s_0=1000\,GeV^2$ is the energy-dependent radius.
With these parameters the proton structure function
$F_2(x,Q^2)$ was calculated \cite{kst} 
using the non-perturbative distribution functions
for the $\bar qq$ component of the photon \cite{kst} and
the dipole cross section (\ref{16}), (\ref{18}). 
The results agree well with the data up to
$Q^2\sim 10\,GeV^2$, which is sufficient for our interval of $k^2$.

The corresponding gluon density (\ref{17})
is shown as function of $k^2$ by the dashed
curve in Fig.~\ref{gg} with the redefined functions 
$\sigma_0(s)$ and $R_0(s)$.
The factor $C(0,s)$ calculated with this gluon density
is shown by thick solid curve as function of 
energy in Fig.~\ref{c(0)}. $C(0,s)$ rises with
energy similar to the prediction of the GBW model.

Summarizing, the two models under consideration provide quite
different values and $k^2$-dependences for the non-integrated
gluon density, as one can see from Fig.~\ref{gg}. Nevertheless,
in the energy range of the E772/E866 experiment at Fermilab
the value $C(0,s) \approx 3 - 4$ is pretty certain and is about
twice as big as the simple two-gluon approximation predicts.

\section{Nuclear shadowing of gluons}\label{gluons}

An important source of broadening
in the transverse momentum of a quark is the gluon radiation
that accompanies the multiple interactions of the quark in the
nucleus. Indeed, one can see from Fig.~\ref{c(0)} that 
the factor $C(0,s)$ is enhanced compared to the
contribution of two-gluon exchange, which corresponds to a
quark scattering on Coulomb gluons without gluon radiation.

It is known that radiation in multiple scattering is 
subject to Landau-Pomeranchuk suppression \cite{lp},
which is a coherence phenomenon in radiation. Namely,
if the gluons radiated due to scattering of a quark
off different nucleons are in phase one should add up the 
amplitudes rather than the probabilities. Interferences
substantially suppress the radiation compared to the classical
expectation (Bethe-Heitler approximation).

The same phenomenon may be treated quite differently as
gluon fusion in
the parton model in the infinite momentum frame of the nucleus, where
it is known as nuclear shadowing for the gluon density at small $x$
\cite{kancheli}-\cite{mv}. Indeed, only the fast part of 
parton clouds of bound nucleons are squeezed by Lorentz contraction,
but the low-$x$ partons may be spread in the longitudinal direction
more than the longitudinal size of the nucleus in its infinite momentum
frame. Therefore, partons originating from different nucleons at the
same impact parameter overlap and may fuse. This effect diminishes 
the number of partons at small $x$.
Thus, we replace the gluon density in (\ref{11})
by a shadowed one, suppressed by a factor $S_A(x,k^2)$ compared
to the unintegrated gluon density in a free proton,
\beq                                                        
C(0,s) = \frac{\pi}{3}\int                          
d^2k\,\frac{\alpha_s(k^2)}{k^2}\,                 
{\cal F}(x,k^2)\,
\,S_A(x,k^2)\ ,              
\label{15a}                                    
\eeq                                           
where $S_A(x,k^2)$ is calculated at $x=4\,k^2/s$. 

\begin{figure}[tbh]
\includegraphics{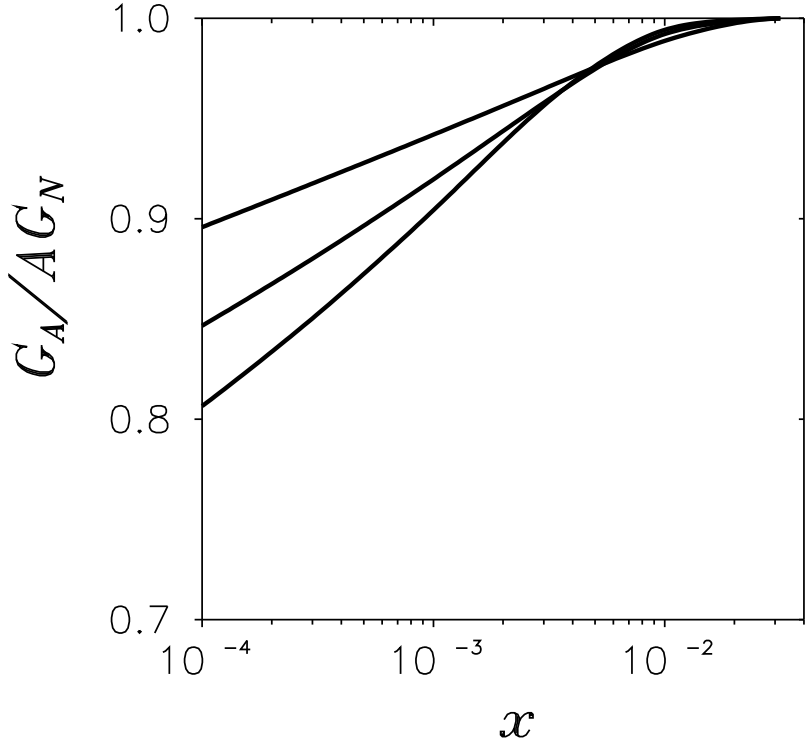}  
\begin{center}
\vspace{10cm}  
\parbox{13cm}
{\caption[Delta]
{\sl The nuclear shadowing factor $S_A(x,k^2)$ for soft gluons
for lead, copper and carbon (from bottom to top)
calculated with the light-cone Green function approach
\cite{kst}.}
\label{soft}}
\end{center}
\end{figure}

The nuclear suppression factor $S_A(x,k^2)$
is calculated in \cite{kst} using the light-cone
Green function approach. At small $k^2$, shadowing
is  controlled by the
strong nonperturbative interactions of gluons, ensuring 
that shadowing is nearly independent of $k^2$ 
for $k^2 < 4\,GeV^2$. This covers the region of $k^2$ we are
interested in, therefore we can safely disregard the $k^2$ dependence
of $S_A(x,k^2)$, except for that which comes from the $x$-dependence.
The factor $S_A(x,k^2)$ is depicted in Fig.~\ref{soft} at small
$k^2$ for a few nuclei as calculated in \cite{kst}. 
We see that the onset of gluon shadowing takes place
at $x < 10^{-2}$, which are smaller values of $x$ than one needs for 
the shadowing of quarks ($F^A_2(x,Q^2)$). Nevertheless, one can see from
Fig.~\ref{gg} that the values of $k^2$ where the unintegrated 
gluon density is large are quite small,
corresponding to very small $x$ where one may expect strong shadowing
effects. 

The results of
calculation of $C(0,s)$ 
for lead with the KST model
for the unintegrated gluon density
are shown in Fig.~\ref{c(0)} by the thin solid curve.
It turns out that the effect of gluon shadowing is not
dramatic, {\it i.e.} does not exceed $10\%$.

\section{The effect of final aperture}\label{aperture}

The factor $C(0,s)$ for the nuclear broadening of the transverse
quark momentum in (\ref{9}) appears as a result of integration
over $k_T$ up to infinity. In reality, the angular
acceptance of the experimental apparatus restricts the accessible
values of $k_T$ to be less than some $k_m$. Introducing this upper
cut off for $k_T$ in (\ref{5a}) we arrive at a modified
value $\tilde C(s)$ that has to replace $C(0,s)$,
\beq
\tilde C(s) = \frac{2\,\pi\,k_{m}}{A\,\la T_A\ra}
\int\limits_0^{\infty}dB\,B\,\int\limits_0^{\infty}
dr\,J_1(k_{m}r)\,F(B,r)\ ,
\label{16a}
\eeq
where
\beqn
F(B,r) &=& \frac{8}{\sigma_{\bar qq}^2(r,s)}\left[
\frac{\partial \sigma_{\bar qq}(r,s)}{\partial r^2}\,
\left(1-\frac{r^2}{3\,\la r_{ch}^2\ra_p}\right) +
\frac{\partial^2 \sigma_{\bar qq}(r,s)}{\partial^2 r^2}\,
r^2\right]\,f_1(B,r)\nonumber\\
&-& \frac{16}{\sigma^3_{\bar qq}(r,s)}\,r^2\,
\left[\frac{\partial \sigma_{\bar qq}(r,s)}
{\partial r^2}\right]^2\,f_2(B,r)\ ,
\label{17a}
\eeqn
and
\beqn
f_1(B,r) &=& 1-(1+a)\,e^{-a}\ ,\\
f_2(B,r) &=& 1-\left(1+a+\frac{a^2}{2}\right)\,e^{-a}\ ,\\                              
a &=& {1\over2}\,\sigma_{\bar qq}(r,s)\,T_A(b)\ .
\label{18a}
\eeqn
 
\begin{figure}[tbh]                                      
\includegraphics{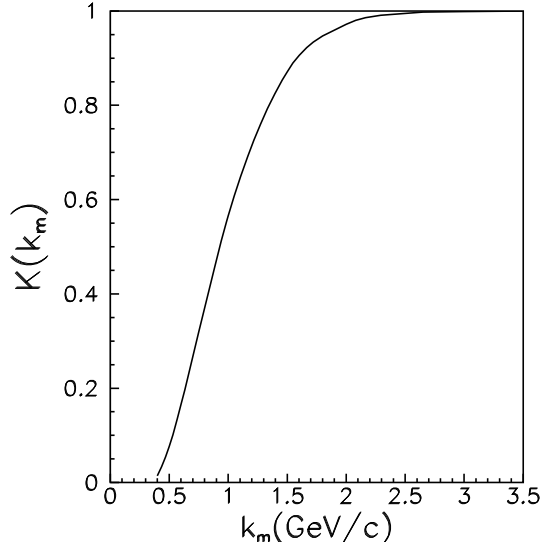}            
\begin{center}        
\vspace{7cm} 
\parbox{13cm}
{\caption[Delta]
{\sl The suppression factor $K(k_m)=\tilde C(s)/C(0,s)$ caused by
a finite aperture calculated with Eq.~\ref{16a} as function of $k_m$.}
\label{k-max}}
\end{center}
\end{figure}

Integration over $B$ and $r$ in (\ref{16a}) needs to be done
numerically. 
The suppression factor $K(k_m)=\tilde C(s)/C(0,s)$
is plotted in Fig.~\ref{k-max} as function of $k_m$.
This substantially deviates from unity for a
cut off $k_m < 2\,GeV$.

\section{Discussion of the results}\label{summary}

We performed calculations for broadening of the transverse momentum of a 
quark propagating through a medium based on the light-cone
color dipole representation, which treats the broadening of transverse
momenta as a color filtering of transverse sizes of $\bar qq$ dipoles 
propagating through nuclear matter. It is natural that
the broadening is proportional to the length of the path times the density 
of the medium, since multiple interactions cause 
the quark to undergo a random walk in the transverse
momentum plane. The main task is to calculate the 
coefficient $C$ which turns out to be the same as in
the dipole cross section $\sigma_{\bar qq}(r_T) = C\,r_T^2$ 
(see Eq.~(\ref{9})).
This is not surprising since, in our approach, the $k_T$ distribution
of the final quarks is proven to have the form of an eikonalized 
dipole cross section. In addition to the formal derivation, we present
another one based on the Moli\`ere scattering theory. 
This simple and intuitive approach helps to understand how the 
propagation of a single quark can be described in terms of the 
color dipole cross section. In particular, it turns out that
a part of the dipole cross section originates from the simple
attenuation of the quark traveling through the medium, while the rest
of the cross section is due to multiple interactions of the quark. 
It is demonstrated that the wide-spread belief that the
lowest order rescattering leads to a broadening 
$\delta\la k_T^2\ra \propto A^{1/3}$ is incorrect.

We have also performed a parameter-free evaluation of the effect of broadening.
We calculated the factor $C(0,s)$ using a phenomenological description
of the dipole cross section that is adjusted to data for the proton structure
function and total hadronic cross sections. We found that the effect of 
$k_T$-broadening steeply rises with energy.

Several corrections diminishing the effect of broadening, such as nuclear
shadowing of gluons and the finiteness of the experimental aperture, 
are evaluated as well.

We make a few remarks in conclusion:
\begin{itemize}
\item
For gluons propagating through a medium, our results
for $k_T$-broadening must be enlarged by the Casimir factor $9/4$.
An example of a process relevant to gluon propagation is
heavy quarkonium production.
\item
Numerically, our results are in a reasonable agreement with
$k_T$-broadening observed in production of $J/\Psi$
and $\Upsilon$ off nuclei, but are somewhat higher than what is
observed in Drell-Yan processes \cite{e772}.
A detailed comparison with available data for Drell-Yan processes,
heavy quarkonium production, dijet and dihadron production off nuclei, etc.,
will be presented elsewhere.
\item
Although we concentrate in this paper on the calculation
of the nuclear broadening of the mean-square
transverse momentum $\delta\la k_T^2\ra$, we are
in a position to calculate the full momentum transfer
distribution of partons after they escape from the nuclear
medium. This distribution is given by Eq.~(\ref{a.12}).
\item
Nuclear broadening of the transverse momentum of partons originating
from relativistic heavy ion collisions is enhanced,
since the parton experiences multiple interactions
propagating through both nuclei. On top of that, if the parton
is produced at mid rapidity with a high transverse momentum
it should substantially increase broadening relative the original 
direction if it propagates through a dense matter ({\it i.e.} 
quark-gluon plasma) since it probes the
density of the medium. This needs measurements with two 
back-to-back hadrons or jets. Thus, the effect of
broadening of $\la k_T^2\ra$ can serve as a sensitive probe
for creation of dense matter in heavy ion collisions.

{\bf Acknowledgements}: We are thankful to J\"org H\"ufner
and Hans-J\"urgen Pirner for helpful discussions. This work was
partially supported  by the Gesellschaft f\"ur
Schwerionenforschung Darmstadt (GSI), grant no. HD H\"UFT.

\end{itemize}


\begin{thebibliography}{99}

\bibitem{e772}  P.L. McGaughey, J.M. Moss and J.C. Peng,
Ann. Rev. Nucl. Part. Sci. {\bf 49} (1999) 217; hep-ph/9905447

\bibitem{hir} B.Z.~Kopeliovich, Soft Component of Hard Reactions
and Nuclear Shadowing (DIS, Drell-Yan reaction, heavy quark production),
in proc. of the Workshop Hirschegg'95: Dynamical
   Properties of Hadrons in Nuclear Matter, Hirschegg, January 16-21,1995,
ed. by H. Feldmeier and W. N\"orenberg, Darmstadt, 1995, p. 102
( hep-ph/9609385)

\bibitem{kst1}  B.Z.~Kopeliovich, A.~Sch\"afer and A.V.~Tarasov,
Phys. Rev. {\bf C59} (1999) 1609

\bibitem{jkt} M.B.~Johnson, B.Z.~Kopeliovich and A.V.~Tarasov,
paper in preparation

\bibitem{emc} J.~Ashman et al., Z.Phys. {\bf C52} (1991) 1

\bibitem{fnal1} M.J.~Corcoran et al., Phys. Lett. {\bf B259} (1991) 209

\bibitem{fnal2} D.~Naples et al., Phys. Rev. Lett. {\bf 72} (1994) 2341

\bibitem{k95} B.Z.~Kopeliovich, Phys. Lett. {\bf B343} (1995) 387

\bibitem{cp} P.~Chiappetta and H.J.~Pirner, Nucl. Phys.
{\bf B291} (1987) 765

\bibitem{hkp} J.~H\"ufner, Y.~Kurihara and H.J.~Pirner,
Phys. Lett. {\bf B215} (1988) 218

\bibitem{zkl} A.B.~Zamolodchikov, B.Z.~Kopeliovich and L.I.~Lapidus,
Sov. Phys. JETP Lett. {\bf 33} (1981) 612

\bibitem{bbgg} G.~Bertch, S.J.~Brodsky, A.S.~Goldhaber and J.R.~Gunion,
Phys. Rev. Lett. {\bf 47} (1981) 267

\bibitem{lr} E.M.~Levin and M.G.~Ryskin, Sov. J. Nucl. Phys. {\bf 33}
(1981) 901; E.M.~Levin, Phys. Lett. {\bf B380} (1996) 399

\bibitem{dhk} J.~Dolej\u{s}i, J.~H\"ufner and B.Z.~Kopeliovich,
 Phys.Lett. {\bf B312} (1993) 235

\bibitem{kz} B.Z.Kopeliovich and B.G.Zakharov,
 Phys.Rev.  {\bf D44} (1991) 3466.

\bibitem{fs} B.~Bl\"attel et al., Phys.  Rev.
 Lett. {\bf 71} (1993) 896

\bibitem{nz94} N.N.~Nikolaev and B.G.~Zakharov,
 Phys. Lett. {\bf B332} (1994) 184

\bibitem{baier} R.~Baier, Yu.L.~Dokshitzer, A.H.~Mueller, S. Peigne and
D. Schiff, Nucl. Phys. {\bf B484} (1997) 265.

\bibitem{wg}  U.~A.~Wiedemann and M.~Gyulassy, Nucl. Phys. {\bf B560} 
(1999) 345

\bibitem{qs} M.~Luo, J.-W.~Qiu and G.~Sterman, Phys. Rev. 
{\bf D50} (1994) 1951

\bibitem{moliere} G.~Moli\`ere, Z. Naturforsch {\bf 2A} (1947) 3;
H.A.~Bethe, Phys. Rev. {\bf 89} (1953) 1256;
W.T.~Scott, Rev. Mod. Phys. {\bf 35} (1963) 231

\bibitem{agk} V.~Abramovsky, V.N.~Gribov and O.V.~Kancheli, 
Sov. J. Nucl. Phys. {\bf 18} (1974) 308

\bibitem{urs}  U.~A.~Wiedemann, {\sl Gluon Radiation off Hard Quarks in a
Nuclear Environment: Opacity Expansion}, hep-ph/0005129

\bibitem{kp1}  B.~Kopeliovich and B.~Povh,
Phys. Lett. {\bf B367} (1996) 329;
Mod. Phys. Lett. {\bf A13} (1998) 3033

\bibitem{kms} J.~Kwiecinski, A.D.~Martin and A.M.~Stasto,
Phys. Rev. {\bf D56} (1997) 3991

\bibitem{pion} S.~Amendolia et al., Nucl. Phys. {\bf B277}
(1986) 186

\bibitem{l} F.E.~Low, Phys. Rev. {\bf D12} (1975) 163

\bibitem{n} S.~Nussinov, Phys. Rev. Lett. {\bf 34} (1975) 1986

\bibitem{garching} T.~Udem et al., Phys. Rev. Lett. {\bf 79}
(1997) 2646

\bibitem{gw}  K.~Golec-Biernat and M.~W\"usthoff,
Phys. Rev. {\bf D59} (1999) 014017;  hep-ph/9903358

\bibitem{kst}  B.Z.~Kopeliovich, A.~Sch\"afer and A.V.~Tarasov,
hep-ph/9908245, to appear in Phys. Rev. {\bf D}

\bibitem{pdt} Review of Particle Physics, R.M.~Barnett et al.,
Phys. Rev. {\bf D54} (1996) 191

\bibitem{lp} L.D.Landau, I.Ya.Pomeranchuk, {\it ZhETF} {\bf 24} (1953) 505, \\
L.D.Landau, I.Ya.Pomeranchuk,  {\it Doklady AN SSSR} {\bf 92} (1953) 535, 735\\
E.L.Feinberg, I.Ya.Pomeranchuk, {\it Doklady AN SSSR} {\bf 93} (1953) 439, \\
I.Ya.Pomeranchuk, {\it Doklady AN SSSR} {\bf 96} (1954) 265, \\
I.Ya.Pomeranchuk, {\it Doklady AN SSSR} {\bf 96} (1954) 481, \\
E.L.Feinberg, I.Ya.Pomeranchuk, {\it Nuovo Cim. Suppl.} {\bf 4} (1956) 652

\bibitem{kancheli} O.V.~Kancheli, Sov. Phys. JETP Lett. {\bf 18}
 (1973) 274

\bibitem{glr} L.V.~Gribov, E.M.~Levin and E.M.~Ryskin, Phys. 
Rept.{\bf 100} (1983) 1

\bibitem{mq} A.H.~Mueller and J.W.~Qiu, Nucl. Phys. {\bf B268} (1986) 427

\bibitem{fls}  L.L.~Frankfurt, S.~Liuti and M.I.~Strikman, Phys. Rev. Lett.
{\bf 65} (1990) 1725

\bibitem{mv} L.~McLerran and R.~Venugopalan, Phys. Rev. {\bf D49} (1994) 2233;
 {\bf D49} (1994) 3352;  {\bf D50} (1994) 2225

\end{thebibliography}
\end{document}